\documentclass[journal=jacsat,manuscript=article]{achemso}

\usepackage[version=3]{mhchem} 

\usepackage{xcolor}

\author{Lucas E. Correa}
\affiliation{Universidade de S\~ao Paulo, Escola de Engenharia de Lorena, DEMAR, 12612-550, Lorena, Brazil}

\author{Pedro P. Ferreira}
\affiliation{Universidade de S\~ao Paulo, Escola de Engenharia de Lorena, DEMAR, 12612-550, Lorena, Brazil}
\alsoaffiliation{Institute of Theoretical and Computational Physics, Graz University of Technology, NAWI Graz, 8010, Graz, Austria}
\email{pedroferreira@usp.br}

\author{Leandro R. de Faria}
\affiliation{Universidade de S\~ao Paulo, Escola de Engenharia de Lorena, DEMAR, 12612-550, Lorena, Brazil}

\author{Vitor M. Fim}
\affiliation{Universidade de S\~ao Paulo, Escola de Engenharia de Lorena, DEMAR, 12612-550, Lorena, Brazil}

\author{Mario S. da Luz}
\affiliation{Instituto de Ciências Tecnológicas e Exatas, Universidade Federal do Triângulo Mineiro, 38025-180, Uberaba, Minas Gerais, Brazil}

\author{Milton S. Torikachvili}
\affiliation{Department of Physics, San Diego State University, San Diego, CA 92182-1233, USA}

\author{Christoph Heil}
\affiliation{Institute of Theoretical and Computational Physics, Graz University of Technology, NAWI Graz, 8010, Graz, Austria}

\author{Luiz T. F. Eleno}
\affiliation{Universidade de S\~ao Paulo, Escola de Engenharia de Lorena, DEMAR, 12612-550, Lorena, Brazil}

\author{Antonio J. S. Machado}
\affiliation{Universidade de S\~ao Paulo, Escola de Engenharia de Lorena, DEMAR, 12612-550, Lorena, Brazil}
\email{ajefferson@usp.br}

\title[]
  {Superconductivity in Te-deficient ZrTe$_2$}

\abbreviations{IR,NMR,UV}

\begin{document}

\begin{tocentry}

\centering
\includegraphics[width=.88\linewidth]{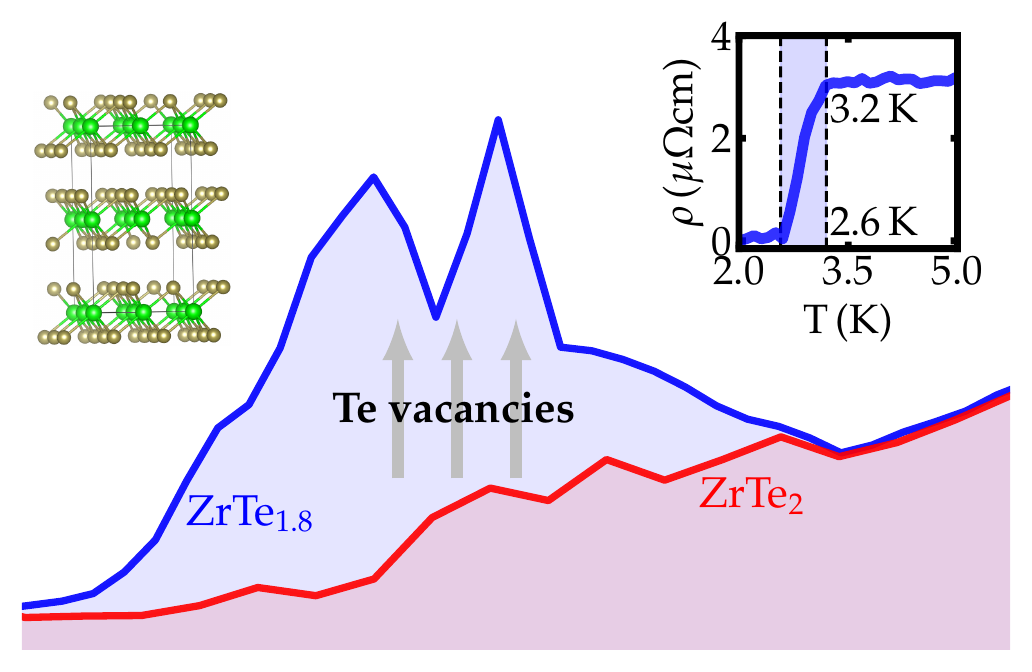}

\end{tocentry}

\begin{abstract}
We present structural, electrical, and thermoelectric potential measurements on high-quality single crystals of ZrTe$_{1.8}$ grown from isothermal chemical vapor transport. These measurements show that the Te-deficient ZrTe$_{1.8}$, which forms the same structure as the non-superconducting ZrTe$_2$, is superconducting below 3.2\,K. The temperature dependence of the upper critical field (H$_{c2}$) deviates from the behavior expected in conventional single-band superconductors, being best described by an electron-phonon two-gap superconducting model with strong intraband coupling. For the ZrTe$_{1.8}$ single crystals, the Seebeck potential measurements suggest that the charge carriers are predominantly negative, in agreement with the ab initio calculations. Through first-principles calculations within DFT, we show that the slight reduction of Te occupancy in ZrTe$_2$ unexpectedly gives origin to density of states peaks at the Fermi level due to the formation of localized Zr-$d$ bands, possibly promoting electronic instabilities at the Fermi level and an increase at the critical temperature according to the standard BCS theory. These findings highlight that the Te deficiency promotes the electronic conditions for the stability of the superconducting ground state, suggesting that defects can fine-tune the electronic structure to support superconductivity.
\end{abstract}

\section{Introduction}
The transition metal dichalcogenides (TMDs), with chemical composition TX$_2$, where T = Zr, Hf, Ti, Mo, W, Ta, etc., and X = S, Se, Te, display a rich number of important physical properties and have the potential for many applications \cite{xu2014,voiry2015,kalantar2015,tedstone2016,choi2017,manzeli2017,han2018,liu2019,zhang2020,wang2020,fu2021}. These materials frequently crystallize in low-dimensional structures and exhibit coherent states and electronic instabilities, such as charge density waves (CDW) and superconductivity (SC) \cite{klemm2015,shi2015,rossnagel2011,zhu2017,hsu2017,heil2017,ying2018,lian2019,balandin2021,xu2021,lian2022}. Their low-dimensional, layered structures are held together by weak van der Waals forces, and SC can emerge upon intercalation of different species in the interstices of the structural van der Waals gap \cite{morosan2006,de2018,dutta2018,sahoo2020}. In addition to SC and CDW, recent studies suggest that many of these TMDs exhibit non-trivial topology, especially type-II topological Dirac states, increasing the interest in this class of materials \cite{belopolski2016,bahramy2018,fei2017,ferreira2021,kar2020}.

The focus of this work is the ZrTe$_{2-x}$ TMD, a Te-deficient compositional modification of the ZrTe$_2$ which crystallizes in the well-known CdI$_2$-prototype structure $P\overline{3}m1$ (164) \cite{jobic1992}, which is attracting intense attention of the physical chemistry research community \cite{zhuang2013, guo2014, rasmussen2015, alyoruk2015, lee2018, wang2019, muhammad2019, ng2020, zhang2020_2, villaos2021, ren2022}. Ionic intercalation can be accomplished in ZrTe$_2$ by positioning the foreign species in the van der Waals gap, frequently leading to SC, e.g. Cu$_x$ZrTe$_2$ (T$_c \approx 9.0$\,K) and Ni$_x$ZrTe$_2$ (T$_c \approx 4.0$\,K) \cite{machado2017,correa2022}. In the case of Ni$_x$ZrTe$_2$, for instance, the possible coexistence of multi-gap superconductivity and CDW (T$_{CDW} \approx 287$\,K) instabilities are well established. Additionally, recent angle-resolved photoemission spectroscopy and de Haas-van Alphen oscillations experiments also suggest that ZrTe$_2$ can be regarded as a Dirac semimetal with massless 4-fold quasiparticles \cite{tsipas2018,nguyen2022}. Due to the possible coexistence of SC and non-trivial topological properties, ZrTe$_2$ is an excellent candidate for probing the interplay between different emergent quantum states.

While the connection between intercalation and SC has been reported in ZrTe$_2$ \cite{machado2017,correa2022}, the effect of slight compositional variations and defects has not been probed. Here we address the effect of Te deficiency on the electronic properties of ZrTe$_2$, and we show that SC can emerge due to slight structural and electronic modifications. Vacancies at the Te sites are introduced in a controlled fashion using isothermal chemical vapor transport (ICVT) growth and the emergence of SC is characterized by means of electrical resistivity, AC susceptibility, and thermoelectric potential measurements. Our measurements suggest that ZrTe$_{1.8}$ is a two-gap superconductor below approximately 3.2\,K with relatively strong intraband coupling. Furthermore, we calculated the band structure of ZrTe$_{1.75}$ within the density functional theory and supercell method. We found that the inclusion of vacancies in bulk ZrTe$_2$ gives rise to a peak at the density of states (DOS) at the Fermi level due to the formation of localized Zr-$d$ bands, bridging the way to electronic instabilities at the Fermi surface.

\section{Methods}

\subsection{Experimental details}

The ZrTe$_{2-x}$ single crystals were prepared by means of the isothermal chemical vapor transport (ICVT), recently proposed by some of the present authors \cite{correa2022_2}. Pre-reacted ZrTe$_{2-x}$ pellets were synthesized by solid-state reaction, and they served as precursors for the ICVT growth. The pellets and a small amount of iodine, which serves as a transport agent, were sealed in a quartz tube, and the growth took place over seven days in temperatures at 950$^\circ$\,C. As a result, the crystals grow out of the pellets, and the typical dimensions of the largest crystals are $10\times10\times0.1$\,mm$^3$. The growth details are discussed in Ref.~\cite{correa2022_2}.

The composition was determined from energy dispersive spectroscopy (EDS) and induced coupling plasma (ICP) utilizing microwave plasma atomic emission spectroscopy (MP-AES) after dilution in HNO$_3$ and HCl. For the MP-AES measurements, sample replicates, reagent blanks, and standard samples with precisely known compositions were used for cross-checking and ensuring accuracy. The crystallographic quality of crystals was verified by X-ray diffraction (XRD) using a Panalytical-Empyrean diffractometer. Rocking curves centered on the $(00l)$ reflections were used to ascertain the orientation and level of crystallinity. 

The electrical resistivity, AC susceptibility, and thermoelectric potential measurements were carried out with the Physical Property Measurement System PPMS-9 from Quantum Design, equipped with a 9.0\,T superconducting magnet. For the 4-probe resistivity measurements, four copper leads were attached to the sample using silver paste. The typical contact resistance was in the 4-5\,$\Omega$ range. The magnetization measurements were performed using the vibration sample magnetometer (VSM). The AC susceptibility measurements were carried out with the ACMS II option of the PPMS, with excitation fields of 1 and 2\,Oe, in the frequency range from 1000-4000\,Hz. The Seebeck coefficient was measured using the thermal transport option of the PPMS-9. The sample was placed across a small printed circuit board section containing four copper lines. Contact of the sample with the copper lines was established with Ni-loaded epoxy.

\subsection{Computational methods}

First-principles electronic-structure calculations were performed within the Kohn-Sham scheme \cite{kohn1965} of the Density Functional Theory (DFT) \cite{hohenberg1964} with scalar-relativistic optimized norm-conserving Vanderbilt pseudopotentials \cite{hamann2013} as implemented in Quantum Espresso \cite{giannozzi2009,giannozzi2017}. Exchange and Correlation (XC) effects were treated with the generalized gradient approximation (GGA) according to vdW-DF2-C09 parametrization \cite{lee2010,cooper2010}, explicitly including the non-local van der Waals interactions. All numerical parameters were exhaustively tested to guarantee a total energy convergence lower than 3\,meV/atom. Based on the convergence results, we adopted a kinetic energy cutoff of 60\,Ry for the wave functions and 240\,Ry for the charge density and potential, and a centered $4\times4\times4$ $k$-point sampling in the first Brillouin according to the Monkhorst-Pack scheme \cite{monkhorst1976}. A denser $24\times24\times24$ $k$-point grid was used to obtain the DOS. Self-consistent-field calculations were carried out using the Methfessel-Paxton smearing \cite{methfessel1989} with a spreading of 0.005\,Ry for Brillouin-zone integration, while the optimized tetrahedron method \cite{kawamura2014} was adopted for the electronic occupation in non-self-consistent-field calculations. All lattice parameters and internal degrees of freedom were relaxed to achieve a ground-state convergence of 10$^{-7}$\,Ry in total energy and 10$^{-6}$\,Ry/$a_0$ for forces acting on the nuclei. The convergence criteria for self-consistency adopted was 10$^{-10}$\,Ry. The supercells were generated with the supercell code \cite{okhotnikov2016}.

\section{Results and discussion}

\begin{figure}[t]
	\includegraphics[width=.6\linewidth]{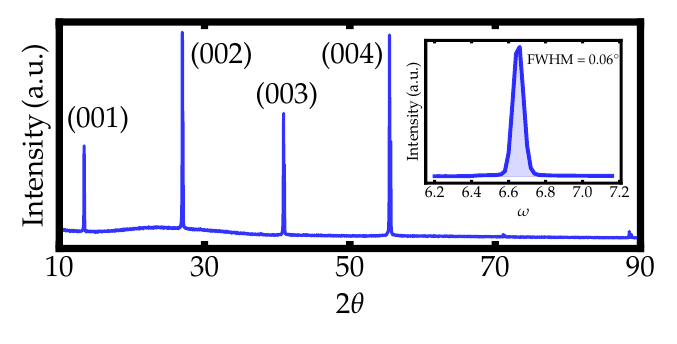}\\
	\includegraphics[width=.3\linewidth]{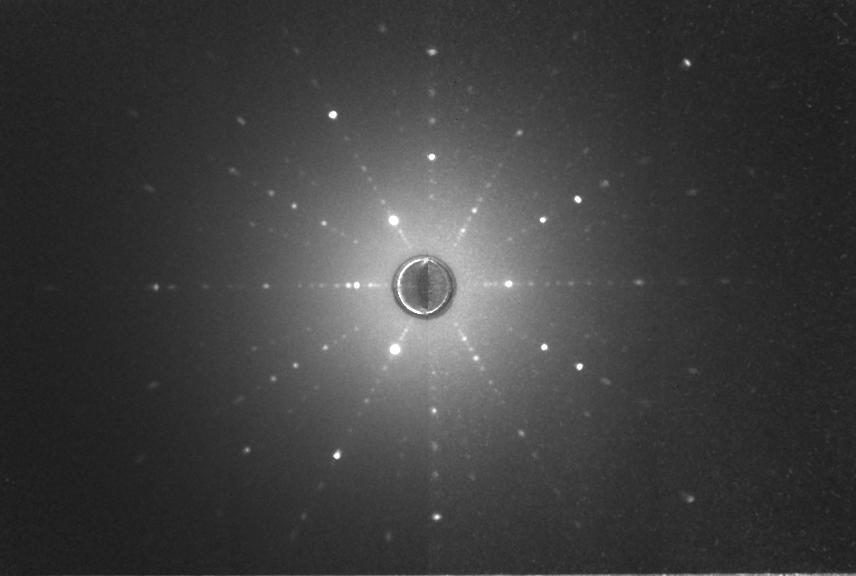}
	\includegraphics[width=.3\linewidth]{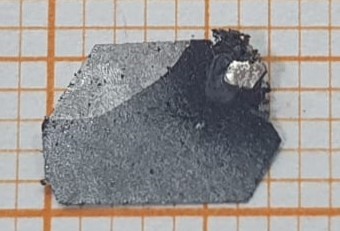}
	\caption{$\theta$-2$\theta$ XRD scan of the ZrTe$_{1.8}$ crystal with beam incident on the flat surface ($ab$-plane). The inset shows the rocking curve centered on the (001) reflection. The lower panel shows the Laue patterns of the ZrTe$_{1.8}$ reciprocal lattice (left) and the picture of one representative single crystal with dimensions $6.5\times5.5\times0.1$\,mm$^3$ (right).}
	\label{fig:drx}
\end{figure}

\begin{figure*}[t]
	\includegraphics[width=.328\linewidth]{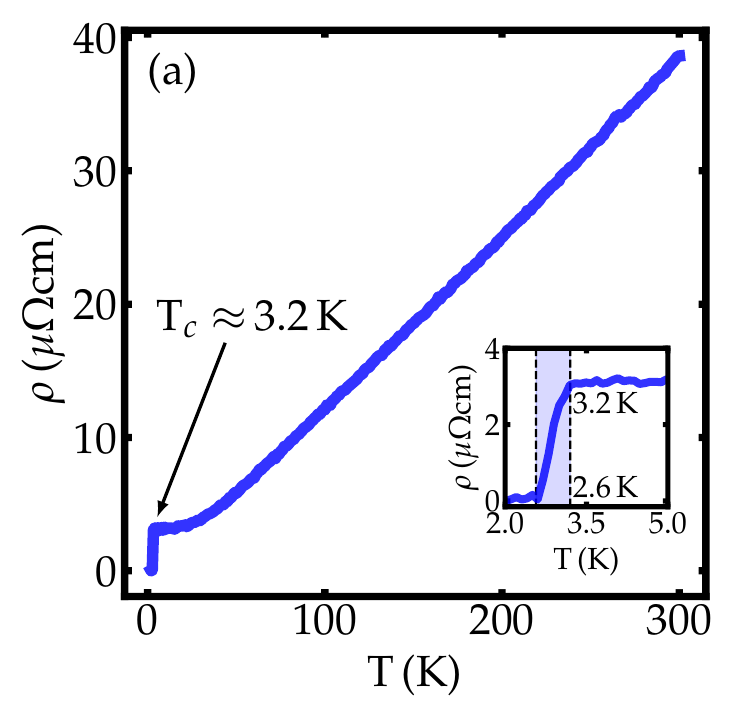}
	\includegraphics[width=.328\linewidth]{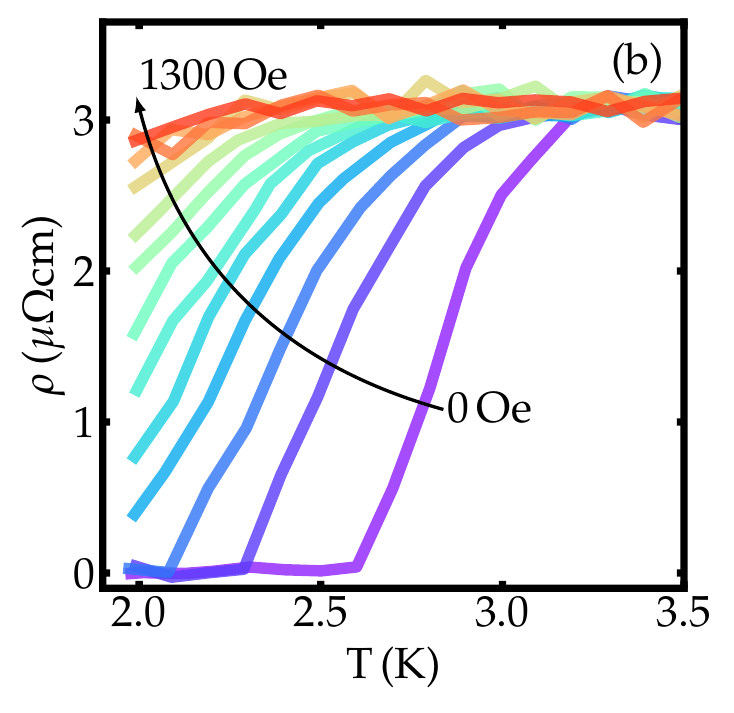}
	\includegraphics[width=.328\linewidth]{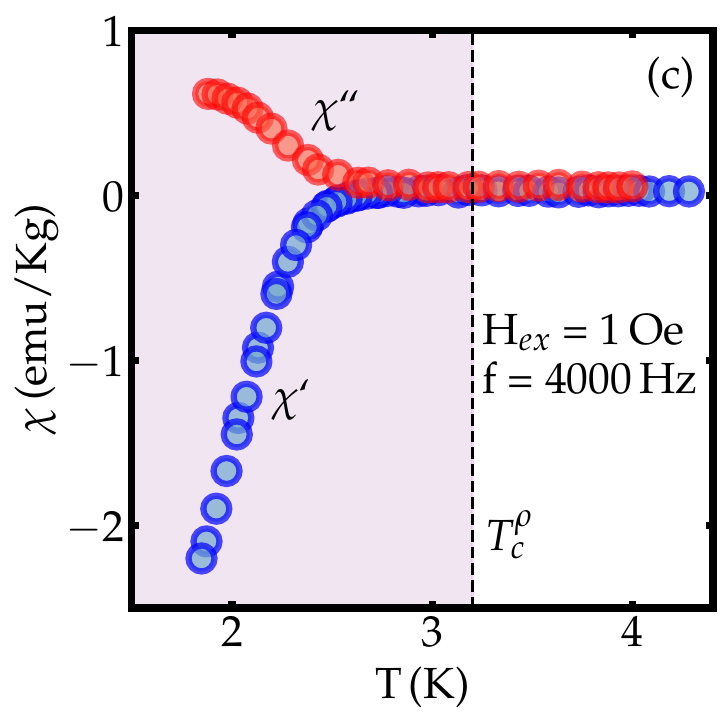}
	\caption{(a) Temperature dependence of the electrical resistivity for the ZrTe$_{1.8}$. The inset presents the low-temperature region, showing the onset of the SC transition near 3.2\,K and the completion at 2.6\,K. (b) Effect of magnetic fields up to 1300\,Oe on the resistive transition to SC. c) AC magnetic susceptibility $\chi$ versus temperature for an applied magnetic field of 1\,Oe and frequency $4000$\,Hz.}
	\label{fig:rho}
\end{figure*}

We synthesized single crystals from a precursor with nominal composition ZrTe$_{1.8}$. The EDS and ICP elemental analysis indicated ZrTe$_{1.8}$ and ZrTe$_{1.85}$ compositions, respectively. For simplicity, heretofore, we will refer to the ZrTe$_{1.8}$ composition. A $\theta$-2$\theta$ XRD scan with the incident beam on the flat surface of a crystal ($ab$-plane) is displayed in Fig.~\ref{fig:drx}. It only shows $(00l)$ reflections, which suggests that flat faces are the basal plane of the trigonal CdI$_2$ structure. The XRD scan is consistent with the Zr--Te phase diagram~\cite{okamoto1999}, where the $P\overline{3}m1$ space group was experimentally determined to be stable in a wide composition range, i.e., ZrTe$_{2-x}$ ($x = 0 – 0.28$). The inset of Fig.~\ref{fig:drx} shows the rocking curve centered on the (001) reflection. This reflection is centered at $\theta = 6.66^\circ$, and the full width at half maximum (FWHM) is $ \approx 0.06^\circ$. The narrow FWHM is consistent with excellent crystallinity. The Te deficient crystals had a dull silvery appearance and typical sizes of $6\times6\times0.05$\,mm$^3$, as shown in Fig.~\ref{fig:drx}.

The temperature dependence of the electrical resistivity of the ZrTe$_{1.8}$ is shown in Fig.~\ref{fig:rho}(a). The resistivity drops nearly linearly upon cooling from 300\,K, it starts leveling off around 40\,K, and eventually drops to zero with the onset of SC at T$_c \approx 3.2$\,K. Fig.~\ref{fig:rho}(b) shows the effect of the magnetic field on the resistive transition to SC in fields up to 1300\,Oe. The shift of the 3.2\,K transition to lower temperature as a function of the applied magnetic field is consistent with SC. Further support for bulk superconductivity is provided by the AC magnetic susceptibility $\chi$ near T$_c$, as shown in Fig.~\ref{fig:rho}c, for data collected with an excitation field H$_{ex} = 1$\,Oe and frequency $f = 4000$\,Hz. While in-phase component $\chi^{'}$ reveals a large diamagnetic signal below T$_c$, the out-of-phase component $\chi^{''}$ increases, reflecting the dissipation associated with the onset of flux dynamics. 

\begin{figure}[h]
	\includegraphics[width=.6\linewidth]{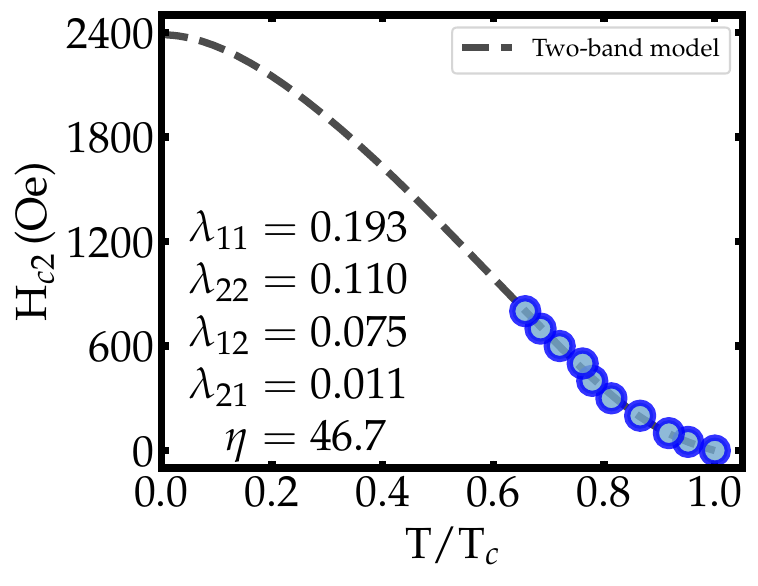}
	\caption{Upper critical magnetic field H$_{c2}$ vs reduced temperature T/T$_c$. The values of H$_{c2}$ were taken from the onset of the resistive transitions. The dashed line is a fit to the data using Gurevich's two-band model.}
	\label{fig:hc2}
\end{figure}

Using the magneto-resistance data of Fig.~\ref{fig:rho} and taking T$_c$ from the onset of the superconducting transitions, the upper critical field H$_{c2}$ can be plotted as a function of the reduced temperature t = T/T$_c$, as shown in Fig.~\ref{fig:hc2}.
The H$_{c2}$(T) data show an upturn below T$_c$, in sharp contrast with the expected quadratic behavior of single-band superconductivity and the Werthamer-Helfand-Hohenberg (WHH) single-gap model \cite{werthamer1966}. The positive curvature of H$_{c2}$ below T$_c$ is frequently taken as an indication of multi-band SC \cite{lyard2002,hunte2008,lei2012,xu2019,bhattacharyya2020,de2021}. Alternatively, the fit of the H$_{c2}$(T) data to the two-band model proposed by Gurevich \cite{gurevich2003} results in an excellent fit, as seen in Fig.~\ref{fig:hc2}, where is represented by the dashed black line. In Gurevich's model \cite{gurevich2003}, the equation for H$_{c2}$ takes the form
\begin{align}
 \label{eq:gurevich}
 a_0&\left[\log t + U(h)\right]\left[\log t + U(\eta h)\right] + a_1\left[\log t + U(h)\right]  \\ &+ a_2\left[\log t + U(\eta h)\right] = 0, \nonumber
\end{align}
where $a_0=2w/\lambda_0$, $a_1=1+\lambda_{-}/\lambda_{0}$, and $a_2=1-\lambda_{-}/\lambda_{0}$, with $\lambda_{-} = \lambda_{11}-\lambda_{22}$, $w = \lambda_{11}\lambda_{22} - \lambda_{12}\lambda_{21}$, and $\lambda_0 = (\lambda_{-}^{2} + 4\lambda_{12}\lambda_{21})^{2}$. The coefficients $\lambda_{mn}$ are the eigenvalues of the BCS superconducting coupling matrix. The diagonal terms $\lambda_{11}$ and $\lambda_{22}$ quantify the intraband coupling, whereas the off-diagonal terms $\lambda_{12}$ and $\lambda_{21}$ describe the interband coupling. The function $U(x)$ is defined as $U(x) = \psi(1/2+x) - \psi(1/2)$, where $\psi(x)$ is the di-gamma function. The arguments of the $U(x)$ in Eq.~\ref{eq:gurevich} are given by $h = H_{c2}D_1/2\phi_0 T$ and $\eta = D_1/D_2$, where $D_1$ and $D_2$ are the intraband electronic diffusivities of bands 1 and 2 in the normal state, and $\phi_0$ is the magnetic flux quanta. The derivation of the intraband diffusivity tensors $D_{m}^{\alpha\beta}$ expressed in terms of microscopic parameters can be found in Appendix A of the Ref.~\cite{gurevich2003}.

The fit to the two-band model yields a H$_{c2}(0)$ value of $\approx 2400$\,Oe, which is consistent with the trend from the H$_{c2}$ data extracted from the resistivity measurements. The diffusivity and coupling parameters yielded by the two-band fit are $\lambda_{11} = 0.193$ and $\lambda_{22} = 0.110$  (intraband coupling), $\lambda_{12} = 0.075$ and $\lambda_{21} = 0.011$  (interband coupling), and $\eta=46.7$. The high diffusivity ratio $\eta$ reflects the significant difference between the electron mobility of distinct Fermi Surface sheets involved in the pairing mechanism, which originates the positive curvature of H$_{c2}(T)$. Still, the $\lambda_{mn}$ values extracted from the effective model suggest that the intraband coupling is one order of magnitude higher than the interband scattering, which likely is the main driving force for the multi-band-type behavior observed. Consistently, SC in ZrTe$_2$ intercalated with Ni and Cu has also been linked to multi-band behavior \cite{machado2017,correa2022}. Given that SC in Te-deficient ZrTe$_{1.8}$ is also consistent with multi-band behavior, we propose that the multi-gap SC state often observed in intercalated TMDs is not necessarily related to the intercalation, but rather it is intrinsically related to the TMDs electronic structure.

\begin{figure}[t]
	\includegraphics[width=.6\linewidth]{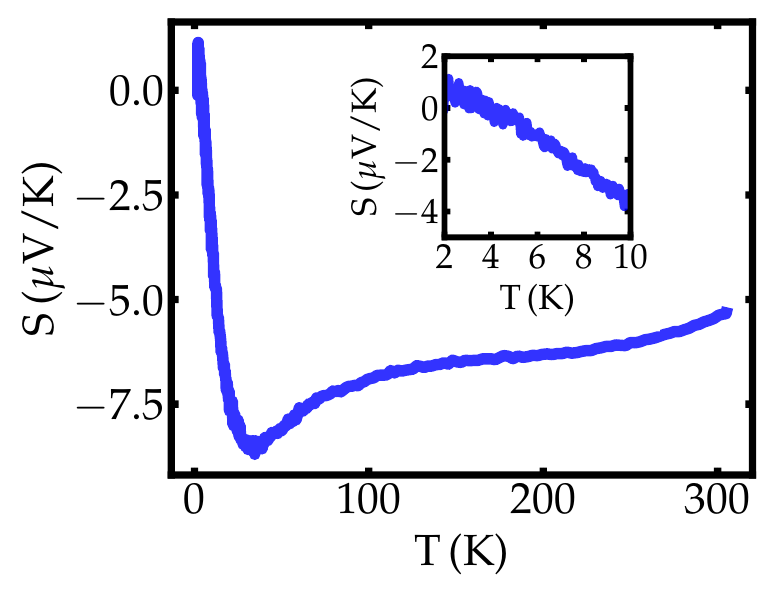}
	\caption{Seebeck coefficient as a function of temperature for ZrTe$_{1.8}$.}
	\label{fig:seeback}
\end{figure}

Measurements of the thermoelectric potential (Fig.~\ref{fig:seeback}) suggest that the preponderance of charge carriers are electrons. The Seebeck potential is $\approx$ -5.2\,$\mu$V/K near ambient temperature. It becomes slightly more negative upon cooling, reaching at minimum ($\approx$ -8.5\,$\mu$V/K) near 35\,K, and increasing rapidly below 20\,K, reaching zero value near T$_c$, consistent with SC pairing. The complex behavior of $S(T)$ near 35\,K possibly results from the convoluted interplay between the temperature dependence of the electron concentration, the asymmetry of the electron distribution near the Fermi level, the mean free path, and mean scattering time. To the best of our knowledge, this is the first time that bulk SC was observed in ZrTe$_2$ , albeit Te-deficient, without intercalation or pressure. The emergence of SC in Te-deficient ZrTe$_2$ suggests that the Te vacancies or the resulting crystalline defects play a crucial role, and a better understanding is still in order.

\begin{figure*}[t]
	\includegraphics[width=.495\linewidth]{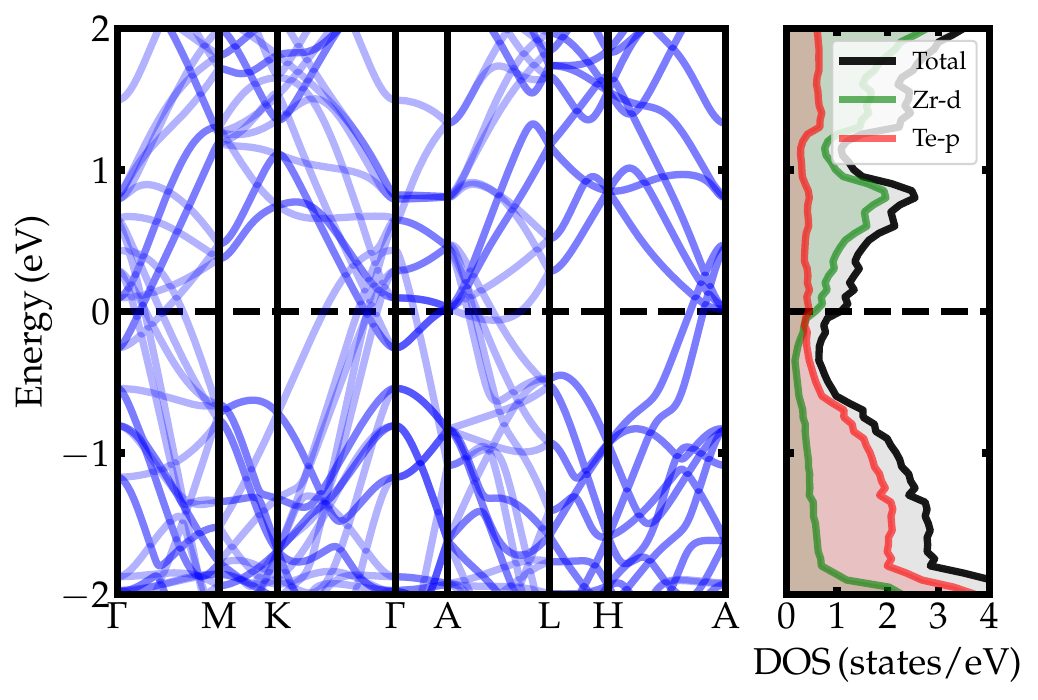}
	\includegraphics[width=.495\linewidth]{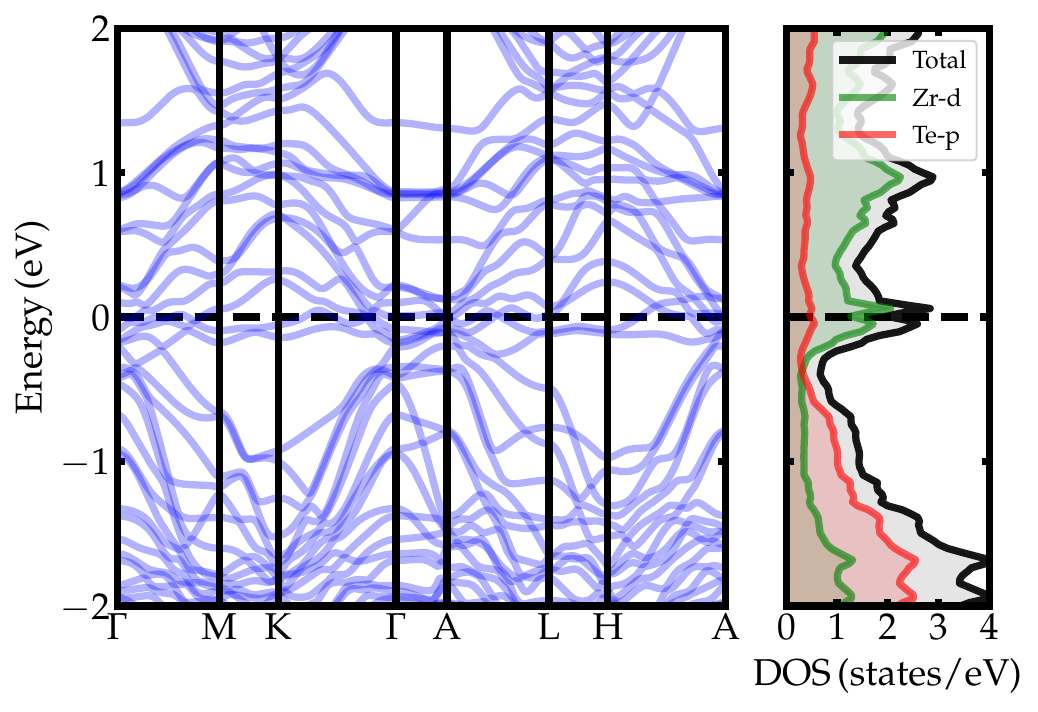}
	\caption{(left) Electronic band structure and DOS for pure ZrTe$_2$ in a $2\times2\times2$ supercell. (right) Electronic band structure for the cluster with the lowest formation energy and averaged DOS for ZrTe$_{1.75}$ in a $2\times2\times2$ supercell. The averaged DOS was weighted according to the degeneracy of each non-equivalent structural configuration.}
	\label{fig:dft}
\end{figure*}

\section{Ab initio assessment}

To probe the effects of tellurium defects, we generated 2$\times$2$\times$2 supercells by occupying 85\,\% of the two non-equivalent Te atomic sites within the $2d$ Wyckoff position of the unit cell. With that, there are 64 possible structural configurations, which we can reduce to only four unique clusters (with different multiplicities) by identifying and employing the full symmetry operations of each supercell. 

Figure \ref{fig:dft} shows the electronic structure for the pure 2$\times$2$\times$2 ZrTe$_2$ supercell and the averaged Te-deficient supercell with composition ZrTe$_{1.75}$ weighted according to the cluster degeneracies. The total DOS at the Fermi level for pure ZrTe$_2$ is 1.1\,states/eV. This value is consistent with previous calculations, which include the spin-orbit coupling \cite{correa2022}, yielding 1.0\,states/eV at the Fermi level, a percentage difference of only 8\,\% with respect to our results without including spin-orbit coupling. From the 1.1\,states/eV, approximately 53\,\% originate from the Zr-$d$ electron pockets, whereas 42\,\% are coming from the Te-p hole pockets. With the inclusion of Te vacancies, a narrow DOS peak at the Fermi level develops due to the formation of localized, near-flat Zr-$d$ electronic bands along the entire extent of the Brillouin zone. The averaged ZrTe$_{1.75}$ has 2.1\,states/eV at the Fermi level, an increase of 92\,\% compared to pure ZrTe$_2$, of which approximately 65\,\% are derived from the Zr-$d$ orbitals, and only 23\,\% are coming from the Te-$p$ orbitals. Since most of the Zr-$d$ states are electron-like pockets at the Fermi surface, the increase of the Zr-$d$ character at the Fermi level is consistent with the polarity of the Seeback potential determined experimentally, revealing that the charge carriers are predominantly negative.

According to the BCS theory, the electronic states that contribute the most to SC are those with energies within a range of the order of $\hbar\omega_D$ around the Fermi energy, where $\omega_D$ is the Debye frequency, and, assuming that the DOS within the energy range $\hbar\omega_D$ is constant, the superconducting critical temperature follows the relation $T_c \approx \hbar\omega_D \exp[-1/N(\epsilon)\lambda]$, where $\lambda$ is the electron-phonon coupling strength \cite{annett2004}. Therefore, the increased DOS at the Fermi level due to the formation of Te vacancies leads to a significant increase in the critical temperature, explaining the relatively high critical temperature in ZrTe$_{1.8}$ compared to the non-superconducting defect-free compound. Furthermore, the Te vacancies also give rise to a sharp DOS peak at E$_F$, similar to van Hove-type singularities, i.e., the DOS varies rapidly within the $\hbar\omega_D$ energy range. When electron correlation effects are considered, this logarithmic instability also enhances T$_c$ and favors spontaneous symmetry-breaking phase transitions \cite{lie1978,pickett1982,mitrovic1983,newns1991,sano2016,chen2020}. Therefore, the appearance of localized states near the Fermi level and the substantial increase of the total DOS at E$_F$ reasonably explains on qualitative grounds the defect-induced SC in ZrTe$_2$.

In order to describe the superconducting state observed in ZrTe$_{1.8}$, one would need to employ advanced techniques such as the fully anisotropic Migdal-Eliashberg theory \cite{margine2013} or Superconducting Density Functional Theory \cite{oliveira1988,luders2005,marques2005} to accurately take into account the anisotropy of the electron-phonon coupling and the two-gap feature of the superconducting gap function. As the presence of Te vacancies necessitates the consideration of supercells and very dense samplings of the Brillouin zone are required to obtain convergence, such a calculation is beyond the scope of the current work, but will be the focus of a future project.

\section{Conclusions}

This work shows the emergence of superconductivity in ZrTe$_{1.8}$, a Te-deficient, off-stoichiometry composition of the non-superconducting TMD ZrTe$_2$, on high-quality single crystals synthesized by ICVT. The superconducting properties were characterized by measurements of electrical resistivity, AC susceptibility, and thermoelectric potential. The ZrTe$_{1.8}$ composition gives rise to a multi-gap superconducting state with a critical temperature close to 3.2\,K. Interestingly, the presence of DOS peak at the Fermi level due to localized Zr-$d$ bands can be linked to the superconducting pairing in the Te-deficient ZrTe$_2$. These results strongly suggest that native point defects, such as vacancies, are essential for SC in the widely investigated class of transition metal dichalcogenides. Furthermore, we show that the multi-band nature is intrinsic to ZrTe$_2$ and that these findings are possibly extending to the whole family of TMDs.

\begin{acknowledgement}

LEC, PPF, LTFE, and AJSM gratefully acknowledge the S\~ao Paulo Research Foundation (FAPESP) under Grants 2018/08819-2, 2019/17878-5, 2019/14359-7, 2020/08258-0, 2021/13441-1, 2021/14322-6. CH acknowledges the Austrian Science Fund (FWF) Project No. P 32144-N36. This study was supported by the Coordena\c c\~ao de Aperfei\c coamento de Pessoal de N\' ivel Superior (CAPES) - Brasil - Finance Code 001, and was conducted by using computational resources of the dCluster of the Graz University of Technology, the VSC-5 of the Vienna University of Technology, and the SDumont supercomputer of the National Laboratory for Scientific Computing (LNCC/MCTI, Brazil).

\end{acknowledgement}

\bibliography{achemso-demo}

\end{document}